\documentclass[pre,floatfix,twocolumn,showpacs,a4]{revtex4}

\usepackage[latin1]{inputenc}
\usepackage[T1]{fontenc}
\usepackage{ae,aecompl}
\usepackage{graphicx}
\usepackage{dcolumn}
\usepackage{amsmath}
\usepackage{amssymb}
\usepackage{tabularx}
\usepackage{epsfig,color}
\usepackage{subfigure}
\usepackage{verbatim}
\usepackage{amsmath, amsthm, amssymb}
\begin{document}

\newcommand{\gs}{$\gamma_s\,$}
\newcommand{\gf}{$\gamma_f\,$}
\newcommand{\kave}{$\left< k \right>\,$ }
\newcommand{\kavefive}{$\left< k \right> = 5\,$ }
\newcommand{\kavenfive}{$\left< k \right> = 5$}
\newcommand{\kaveten}{$\left< k \right> = 10\,$ }
\newcommand{\cave}{$\left< c \right>\,$ }
\newcommand{\er}{Erd\H{o}s-R\'enyi } 
\newcommand{\kaven}{$\left< k \right>$} 
\newcommand{\kcout}{$k_{c,out}$ }
\newcommand{\kcoutn}{$k_{c,out}$}    
\newcommand{\kcoutave}{$\left< k_{c,out} \right> $ }
\newcommand{\knin}{$k_{n,in}$ }
\newcommand{\kninave}{$\left< k_{n,in} \right> $ }
\newcommand{\knout}{$kn_{out}$ }
\newcommand{\knoutave}{$\left< k_{n,out} \right> $ }
\newcommand{\knoutavedef}{$\left< k_{n,out} \right> = k_{c,out} / s $ }
\newcommand{\rdef}{ $ r = k_{n,in} / \left<   k_{n,out} \right> $ }
\newcommand{\rdefn}{ $ r = k_{n,in} / \left<   k_{n,out} \right> $}
\newcommand{\rdefs}{ $ r = s(s-1) /  k_{c,out}  $ }
\newcommand{\rdefsn}{ $ r = s(s-1) /  k_{c,out}$}
\newcommand{\rdeflong}{ $r = k_{n,in} / \left<   k_{n,out} \right>   = (s-1) / (  k_{c,out} / s) =  s(s-1) /  k_{c,out} $}
\newcommand{\fmin}{$f_{min}$ }
\newcommand{\ERhom}{ERH }
\newcommand{\ERhomn}{ERH}
\newcommand{\pERhom}{PERH }
\newcommand{\pERhomn}{PERH}
\newcommand{\EDhom}{EDH }
\newcommand{\EDhomn}{EDH}
\newcommand{\EDexp}{EDE }
\newcommand{\EDexpn}{EDE}

\title[Short Title]{Broad lifetime distributions for ordering dynamics in complex networks}
\author{R. Toivonen$^1$}\email{Riitta.Toivonen@gmail.com}
\author{X. Castell\'o$^2$}
\author{V. M. Egu\'iluz$^2$}
\author{J. Saram\"{a}ki$^1$}
\author{K. Kaski$^1$}
\author{M. San Miguel$^{1,2}$}
\affiliation{$^1$Department of Biomedical Engineering and Computational Science (BECS), Helsinki University of Technology, P.O. Box 9203,
FIN-02015 HUT, Finland \\ $^2$IFISC(CSIC-UIB), Institute for Cross-Disciplinary Physics and Complex Systems, Campus Universitat Illes Balears,
E-07122 Palma de Mallorca, Spain}

\date{\today}
\begin{abstract}

We search for conditions under which a characteristic time scale for
ordering dynamics towards either of two absorbing states in a finite
complex network of interactions does not exist.  With this aim, we study random
networks and networks with mesoscale community structure built up from
randomly connected cliques. We find that large heterogeneity at the
mesoscale level of the network appears to be a sufficient mechanism
for the absence of a characteristic time for the dynamics. Such
heterogeneity results in dynamical metastable states that survive at
any time scale.

\end{abstract}


\maketitle

\section{\label{sec:intro} Introduction}

A key characteristic for nonequilibrium dynamics of interacting many
body systems is the relaxation time. Typically, finite systems starting
from generic initial conditions far from equilibrium reach a final
stationary state or attractor in a characteristic time. For simple
nonequilibrium lattice models~\cite{Marro_Dickman_1999} the dynamics
often leads to an absorbing state. In some cases the system might get
trapped in a metastable state, which generally also has a well defined
expected lifetime. Frozen metastable configurations that persist
indefinitely in the absence of fluctuations are also possible. An
intriguing situation occurs when such a characteristic time cannot be
defined. In particular this has been shown to occur when the system
reaches different dynamical (nonfrozen) metastable states with very
different lifetimes~\cite{comm_epl}.

The two basic inputs in a model of interacting units are the
interaction rules among units, and the network of interactions,
i.e., links in the network defining who interacts with whom. The
structure of this network is expected to strongly influence the
dynamics and collective behavior arising from interactions.  A
particular feature of networks of social interaction is that they are
structured into cohesive groups within which the internal links are
dense, and which are sparsely
interconnected~\cite{scott,SantoReview,newmanCommunitiesReview2004,ComparingCommunityDetection2005,CommunityIdentificationReview2007,Radicchi_comms,Arenas_comms,GirvanNewman,kclique,Palla:2007qy,Kumpula:2007ly}.
Such \emph{communities} are known to have a deep
impact on the dynamics taking place in the network. For example, for
oscillators coupled via a complex network, synchronization takes place
first within highly interconnected local structures, and synchronized
domains expand via intercommunity
connections~\cite{oscillatorsAndMotifs,synchronizationAndModularity,synchronizationModularityAndHierarchy,synchronizationIntraModuleConnections,synchronizationInCommunities,SynchronizationReview}.
Similarly, information has been shown to spread rapidly within
communities, but slowly across the network, particularly if
intercommunity links are weak~\cite{mobile1}. Communities may also
promote the emergence and survival of cooperation~\cite{lozano}.

In dynamics of competing options, research has focused on the question
of whether and how a system-wide consensus is
reached~\cite{SantoReview}.  For a two-spin system following the
majority rule~\cite{MajorityRuleRedner}, network topologies with
communities can be constructed in which no consensus will take
place~\cite{Lambiotte:2007oq}. For a three-state model of competing
options~\cite{Bilingual1}, the absence of a characteristic time until
consensus is reached~\cite{comm_epl} seems to be associated with
metastable traps caused by community structure.  Dynamics sensitive to
community structure have also been employed for identifying
communities, including various spin systems such as in the Ising
model~\cite{isingInCommunities,IsingInGraphPartitioning}, the Potts
model~\cite{ReichardtBornholdt2004,SuperparamagneticClustering}, and
models of random
walks~\cite{RandomWalksInCommunityDetection,communitiesWithRandomWalks,communitiesWithWeightedRandomWalks}.

The question of general conditions under which a broad distribution of
relaxation times is obtained merits a detailed and systematic
study. In this paper we address this question by investigating the
role of the network of interactions, in particular its mesoscopic
structure and topological traps, in nonequilibrium ordering
dynamics. For this, we will consider the AB model~\cite{Bilingual1}
which describes ordering dynamics towards an absorbing $A$ or $B$
state. This model is an extension of the simple two-state voter
model~\cite{Holley:1975}, in which dynamics at the interfaces has been
proven to evolve by curvature reduction~\cite{Bilingual1} in contrast
to the noisy interface dynamics of the voter
model~\cite{Dornic2001,Al_Hammal_Chate_2005}. Thus, the results
reported in this paper might be generic for a class of models in which
the dynamics at the interfaces is curvature driven, such as the spin-flip kinetic Ising model (SFKI)~\cite{Gunton1983}. In addition,
mesoscopic community structure of the networks is especially relevant
in discussions of the AB model which was originally motivated by
studies of language
competition~\cite{ASmodel,Minett_2008,Castello2008,Bilingual1}. We
focus on the question of which features of the network topology, such
as relatively isolated groups of nodes, the presence of communities,
their interconnectivity, or their size distribution, give rise to
trapped metastable states. Here we study the dynamics of the AB model
in a controlled setting by constructing test case networks in which
community boundaries are clear.

The outline of the paper is as follows: In Sec.~\ref{sec:ABmodel}
we introduce and briefly review the AB model. We first consider random
networks (Sec.~\ref{sec:ER}) where we identify substructures
causing broad distributions of lifetimes in random networks with low
mean degree. To explain the observations therein, we introduce
the concept of {\em dynamical robustness} measured by the survival
time (i.e., a characteristic relaxation time) of relatively isolated groups of
nodes in a state different from the one in the final absorbing
state. In Sec.~\ref{sec:communities} we study the effect of the
presence of communities starting from an underlying random network and
communities of equal size, and considering later an exponential
distribution for the sizes of different communities. The results and
conclusions are summarized in Sec.~\ref{sec:discussion}.

\section{The AB model}
\label{sec:ABmodel}

The AB model~\cite{Bilingual1} considers a set of interacting units
located at the nodes of a network. Each unit can be in either state $A$
or $B$, or in a third state $AB$ of coexisting options at the individual
level. Thus the model describes the competition of two nonexcluding
options or states. It can be seen as an extension of the two-state
voter model~\cite{Holley:1975} by the introduction of a third
nonsymmetric mixed state.  A node changes its state with a
probability that depends on the states of its neighbors in the
network. The fraction of neighbors of an agent in each state ($A$, $B$,
$AB$) are called the \emph{local densities}: $\sigma_A$, $\sigma_{B}$,
and $\sigma_{AB}$. Transitions from $A$ to $B$ always take place via the
third mixed $AB$ state. The AB model is defined by the following
transition probabilities:
\begin{eqnarray}
p_{A\to AB} = \frac{1}{2} \sigma_B, & & \; p_{B\to AB} =
\frac{1}{2} \sigma_A \label{eq:ABa} \\
p_{AB \to A} = \frac{1}{2} (1-\sigma_B), && p_{AB \to B} = \frac{1}{2} ( 1- \sigma_A)~.
\label{eq:ABb}
\end{eqnarray}

The mean-field analysis of these equations shows that there exist
three fixed points with two of them stable and one unstable. The
stable fixed points correspond to consensus in either of the two
options $A$ or $B$, while the unstable one corresponds to a situation
where the three states $A$, $B$ and $AB$ coexist.  In our simulations we
initially set each node in a network of size $N$ randomly to one of
the states A, B, or AB. At each time step all nodes are updated in
random order according to the transition probabilities
Eqs.~(\ref{eq:ABa}) and (\ref{eq:ABb}). The {\em lifetime} of a run is
defined as the number of time steps it takes to reach either of the
absorbing states. We monitor the {\em fraction of alive runs}, that
is, the fraction of simulations which still have not reached an
absorbing state at time {\it t}, $P(t)=1-\int_0^{t} p(t')dt'$, where
$p(t)$ is the probability distribution of lifetimes.  In a
two-dimensional regular and finite lattice, the system reaches in a
finite time and with probability one an absorbing state,
i.e., consensus in either of the single-option states, $A$ or $B$ (with the
same probability~\cite{Bilingual1}). Trapped metastable states have
been observed in the AB model dynamics in a two-dimensional lattice, in
which they take a stripelike form~\cite{Bilingual1}, but with
well-defined mean lifetime (lifetimes of these metastable states are
distributed exponentially).

In contrast with the voter model, the AB model dynamics is curvature
driven~\cite{Bilingual1}, such that the $A$ and $B$ domains form and grow
in size by a coarsening process in which the characteristic length of
a domain, averaged over different realizations of the dynamics, grows
as $\left<\xi(t)\right>\sim t^{\gamma}$, $\gamma\simeq 0.5$.  The $AB$
agents never form domains but instead place themselves in the
boundaries between $A$ and $B$ domains, transforming the noisy interface
dynamics of the voter model into a curvature driven dynamics.

In a network with community structure~\cite{TOSHK}, the 
boundaries between homogeneous single-option domains tend to follow
the community boundaries~\cite{comm_epl}. The most interesting result
for the AB model in these networks is the fact that there is no
characteristic time scale for the dynamics: a power law distribution
for the distribution of lifetimes is obtained with an exponent such
that a characteristic time cannot be defined. This behavior seems to
be associated with the existence of dynamical long-lived trapped
metastable states that survive at any time scale. In order to obtain a
fuller understanding of this phenomenon, in the remaining part of the
paper we study the $AB$ dynamics in networks with a predesigned
mesoscopic structure.

\begin{figure}[tb]
\includegraphics[width=1.0\linewidth]{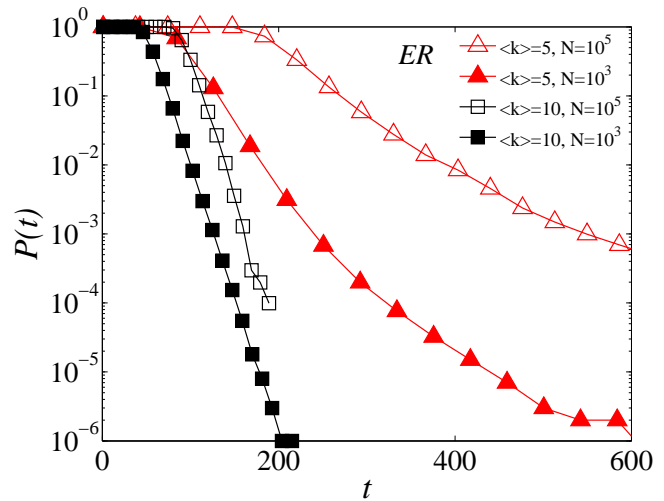}
\caption{(Color online) Fraction $P(t)$ of runs alive at time $t$
  for \er (ER) networks with different average degrees ({$\Box$}:
  \kaven$=10$, \mbox{\textcolor{red}{$\triangle$}: \kaven$=5$}). Solid
  symbols, network size $N=10^3$; open symbols, $N=10^5$. Averaged
  over 100 network realizations, 100 runs in each (1000 in each for
  $N=10^3$).  }
\label{fig:ER}
\end{figure}

\section{The AB model in random networks}
\label{sec:ER}

\subsection{\er Networks}
\label{sec:ER-nets}

Here we first consider the \er (ER) random network topology~\cite{ER},
in which each of the possible links between the $N$ nodes is present
with probability $p$. The network can equivalently be characterized by
the average degree \kaven$=p(N-1)$. We find that in \er
networks the fraction of alive runs $P(t)$ depends unexpectedly on
link density [Fig.~\ref{fig:ER}], such that, for high link densities up
to a fully connected network, $P(t)$ is clearly exponential, but for
low link densities it turns out to be broader, indicating the
existence of metastable states. The cases of \kave$=10$ and \kave$=5$
are selected to illustrate this difference.

\begin{figure}[tb]
\centering
\includegraphics[width=1.0\linewidth]{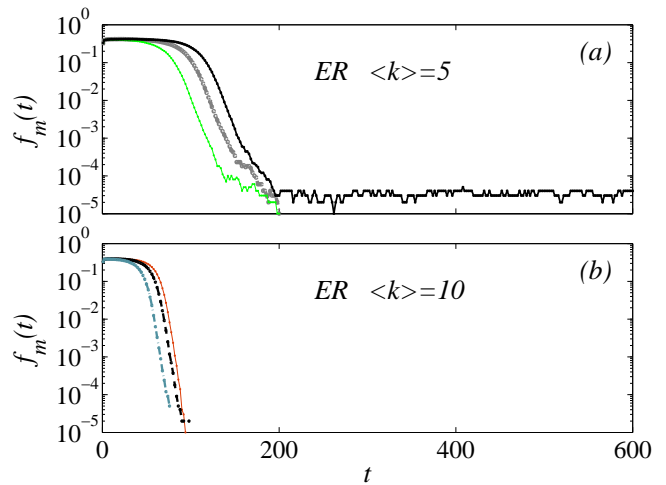}
\caption{(Color online) Time evolution of the fraction $f_{m}$ of
nodes in the state ($A$ or $B$) that became the minority and finally died
out, in several individual runs. (a) ER \kave$=5$, $N=10^5$; (b) ER
\mbox{\kave$=10$}, $N=10^5$. }
\label{fig:metastable}
\end{figure}

\begin{figure}[tb]
\includegraphics[width=0.6\linewidth]{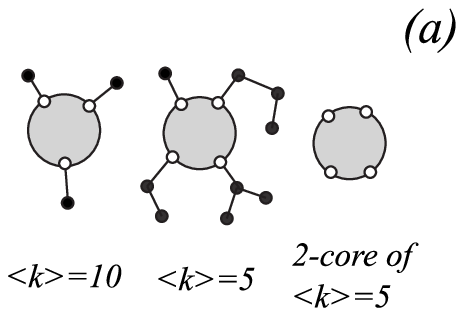}
\includegraphics[width=1.0\linewidth]{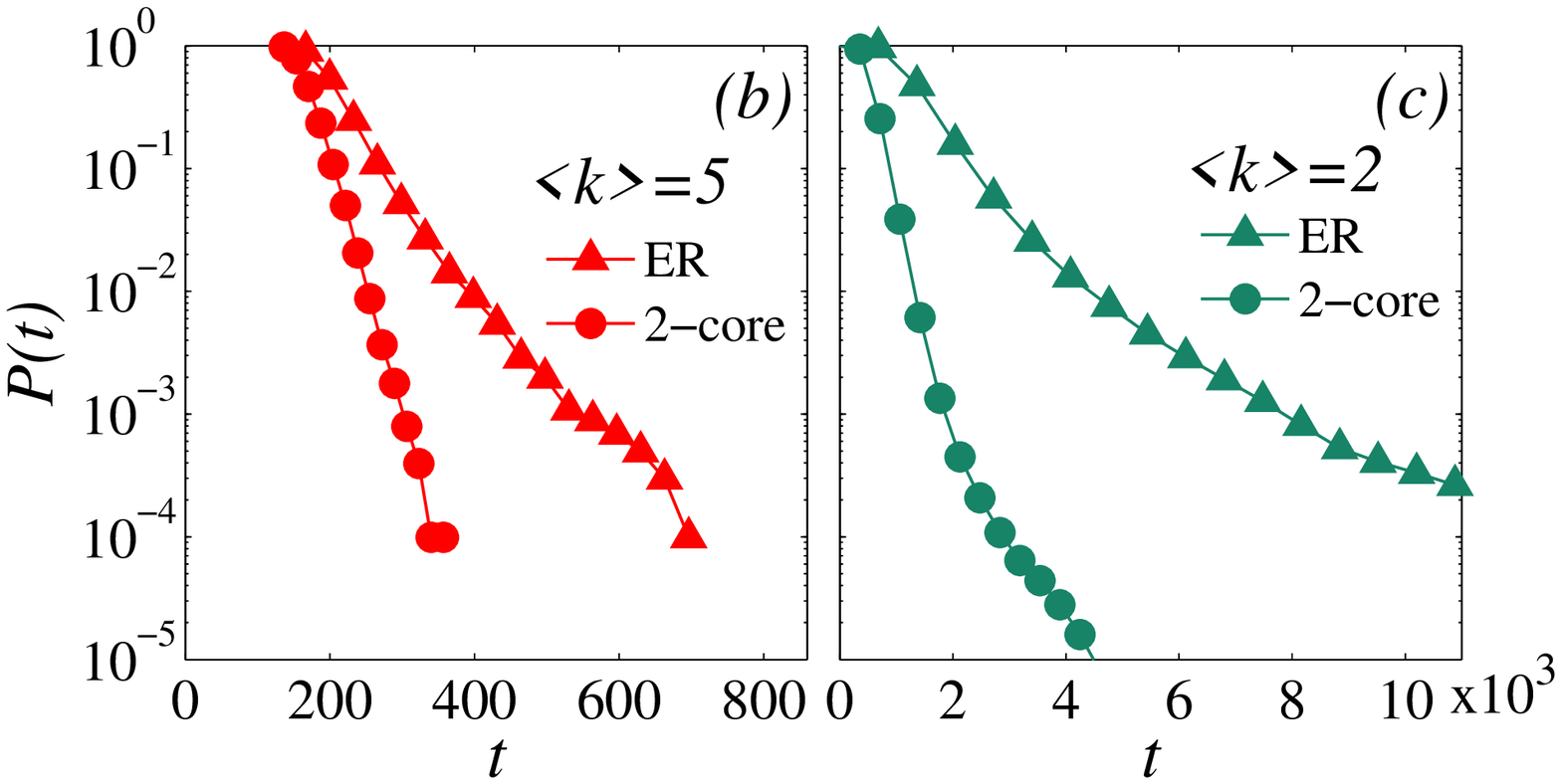}
\centering
\caption{(Color online) (a) Schematic illustration of branches in
the different ER networks studied. The root nodes are indicated by
open circles. (b) The fraction of alive runs $P(t)$ for ER networks
with \kave$=5$ and $N=10^5$, and their two-cores. (c) ER networks with
\kaven$=2$ and $N=2\times10^3$, and their two-cores.  Averaged over 100
network realizations, 100 runs in each (5000 network realizations for
two-core of \kaven$=2$). }\label{fig:2core}
\end{figure}

The metastable states are visualized by displaying the fraction $f_m$
of nodes in the minority state (the state $A$ or $B$ that becomes the
minority and finally dies out) in individual runs
[Fig.~\ref{fig:metastable}(a)]. Figure~\ref{fig:metastable}(b) shows
that metastable states do not arise for \kave$=10$.  The observed
metastable states concern only a very small fraction of nodes.
Further analysis reveals that they are associated primarily with
``branches''. We use the term branch here to indicate maximal treelike
substructures that are connected to the rest of the network through a
node that has degree $k>2$, which we will call here the root node of
the branch [see Fig.~\ref{fig:2core}(a) for illustration].  Branches
can be removed from the network (except for the root node) by
successively removing nodes of degree $k=1$ until all the remaining
nodes have degree $k\ge2$, i.e., by taking the {\em two-core} of the
network~\cite{costaMeasurements}. We call the maximal network distance
from the root node to any other nodes in the branch the diameter of
the branch. Figure~\ref{fig:2core}(a) displays schematically the
difference between the ER networks with \mbox{\kave$=5$} and
$10$ with respect to the presence of branches. In the ER
networks with \kaveten typically only branches with unit diameter are
present, while branches with diameter $2$ or $3$ arise frequently in ER
nets with \mbox{\kave$=5$}, often with bifurcations.

Taking the two-core of ER networks with \kave$=5$ and running the AB
model in the resulting networks confirms the role of branches in
producing a deviation from an exponential distribution in $P(t)$, see
Fig.~\ref{fig:2core}(b). The metastable states disappear and an
exponential lifetime distribution is recovered.  In very sparse ER
networks, illustrated by the case \kaven$=2$ in Fig.~\ref{fig:2core}(c), a
slight deviation not explained by branches remains that could be
attributed to the largely chainlike structure of the
networks~\footnote{Note also that it has been argued that modularity
may arise from fluctuations in sparse ER
networks~\cite{ModularityFromFluctuations}}. However, the branches are
clearly responsible for the longest-lived metastable states.

\subsection{Dynamical robustness and survival times}
\label{sec:resitance_against_invasions}

\begin{figure}[htb]
\centering
\includegraphics[width=0.8\linewidth]{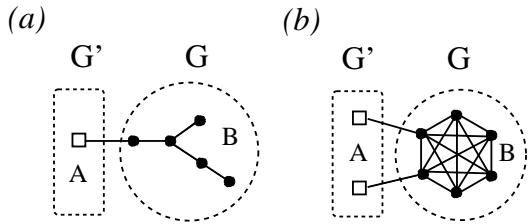}
\caption{Characterization of the dynamical robustness against invasion
  for a given topological structure {\it G}. We show a schematic view
  for two example cases: (a) a branch excluding the root node and (b)
  a clique.}
\label{schematic:dynamical robustness}
\end{figure}

In order to characterize the behavior of relatively isolated groups of
nodes that remain in the minority state after most of the network has
homogenized in either state $A$ or $B$, we introduce the concept of {\em
dynamical robustness} against invasion for a given topological
structure. It concerns a group of nodes $G$ subjected to dynamics of
competing options. It measures the resistance of $G$ against consenting
to outside pressure applied to $G$ by its neighbors $G'$
[Fig~\ref{schematic:dynamical robustness}]. We initialize the nodes in
$G$ to state $B$, and fix the nodes in $G'$ permanently to state $A$. The
dynamical robustness of $G$ is then characterized by a survival time,
$\tau$, defined as the characteristic time it takes for $G$ to
homogenize to state $A$. A substructure $G$ is dynamically robust when it
has a large survival time.

\begin{figure}[tb]
\includegraphics[width=1.0\linewidth]{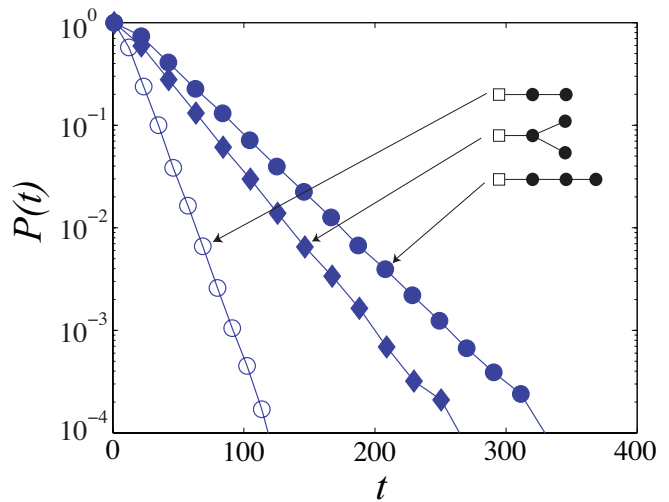}
\centering
\caption{(Color online) Fraction of alive runs $P(t)$ for a chain (a
  branch with no bifurcations) with diameter $2$ (open circles), a
  branch with diameter $2$ with a single bifurcation (diamonds), and a
  chain with diameter $3$ (solid circles), initialized such that the
  nodes denoted by open squares are permanently set to one state, and
  the remaining nodes are initially set to the opposing state. We
  performed $10\,000$ runs in each topology.}\label{fig:chains}
\end{figure}

As an example, consider the dynamical robustness of branches with
diameters $2$ and $3$, one of them containing a bifurcation
[Fig. \ref{fig:chains}]. The time it takes for each of these
topologies to homogenize to the consensus state appears to be
distributed exponentially, $P(t)\sim e^{-t/\tau}$ but with a different
characteristic time $\tau$, corresponding to different time scales. For a
branch with given diameter, bifurcations increase the survival
time. It is noteworthy that, compared to the times it takes for an \er
network with no branches to reach consensus [Fig. \ref{fig:2core}(b)],
the time that a single branch may remain with the minority opinion is
very long.

Each branch of different diameter, and with a different number of
bifurcations, produces an exponential lifetime distribution with a
different characteristic time $\tau$. The combination of various time scales
leads to the observed broader than exponential lifetime distribution
in low-link-density ER networks.

\section{The role of communities}
\label{sec:communities}

\subsection{Networks with equally sized cliques}
\label{sec:ERhom}

In this section, we discuss the effect of communities on lifetimes of
the system, using simple test networks with equally sized
communities. To keep things as clear as possible, we use cliques,
i.e., fully connected graphs, as communities.  We denote by \kcout the
clique out-degree, or the number of links connecting each clique to
outside nodes (note that the term out-degree here does not refer to
directed edges). For a node in a clique of size $s$, we denote the
number of links to the other nodes within the clique by \knin$=s-1$,
and the average number of links to nodes outside the clique by
\knoutavedef.

\begin{figure}[htb]
\centering
 \includegraphics[width=1.0\linewidth]{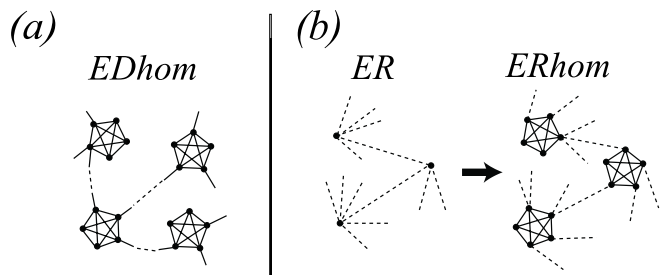}
\caption{Generation of networks with equally sized cliques: (a)
\EDhomn and (b) \ERhomn. To obtain an \ERhom network of size $N$, we
begin with an underlying ER network with $N/s$ nodes, where $s$ is the
clique size. }
\label{schematic:ERhom}
\end{figure}

We employ two different methods for connecting the cliques. In the
first construction, \kcout is equal for all cliques. We label these
networks EDH, for equal out-degree and homogeneously sized
cliques. The \kcout link ends are assigned to randomly selected nodes
in each clique, and then randomly paired, under the condition that two
link ends from the same clique are not allowed to be connected
[Fig.~\ref{schematic:ERhom}(a)]. In the second construction, we begin
with an ER network, with high link-density to avoid branches with
diameter larger than $1$, and replace its nodes with equally sized
cliques (Fig.~\ref{schematic:ERhom}(b)). Each link of the underlying
\er network, connecting two cliques, is again assigned to a uniformly
randomly selected node in each clique.  We label these networks ERH,
as they are based on an underlying ER network, and consist of
homogeneously sized cliques. In \ERhom networks, \kcout is distributed
according to the Poisson distribution of the underlying ER network.

\begin{figure}[htb]
\centering
\includegraphics[width=1.0\linewidth]{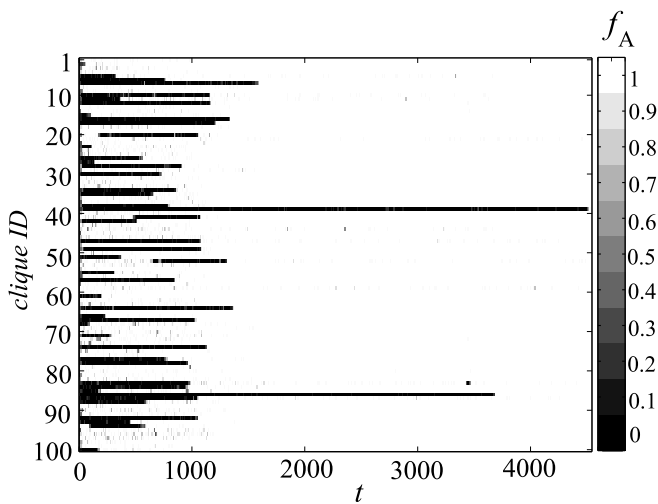}
\caption{Time evolution of the fraction of $A$ agents $f_A$ in each
clique, labeled with IDs from $1$ to $100$, in a single run in an
\ERhom network with clique size $s=10$, $N=1000$, and
\kcoutave$=10$. The resolution is 10 time steps.}
\label{fig:communityStates_ERhom_s10}
\end{figure}

We run the AB model dynamics in different realizations of the \EDhom
and \ERhom networks with \mbox{\kcoutave$=10$} and various clique
sizes $s$, always starting from random initial conditions.  Let us
first obtain a detailed view of the time evolution of the dynamics by
monitoring the fraction of agents in each state within each
clique. Note that, because the $AB$ agents do not tend to form
$AB$ domains, the densities $f_A$ and $f_B$ of $A$ and $B$ agents within
each clique will be practically complementary ($f_B\approx1-f_A$).
Figure~\ref{fig:communityStates_ERhom_s10} displays $f_A$ within each
clique for a run in an \ERhom network, indicated by gray scale from
white (all A agents) to black (no A agents).  Each row corresponds to
one clique. The randomly initialized cliques very rapidly homogenize
to either state $A$ (white) or $B$ (black).  The plot shows that cliques
remain homogenized to either state $A$ or $B$ during most of the run, and
that they do not often flip from one state to the other (this is also
true for the \EDhom networks, not shown). Two of the clusters remained
in the minority state $B$ for long after the rest of the network was
homogenized to the opposing state, indicating a metastable
state. These appeared frequently in the \ERhom networks, in contrast
to the \EDhom networks.

\begin{figure}[htb]
\centering
\includegraphics[width=1.0\linewidth]{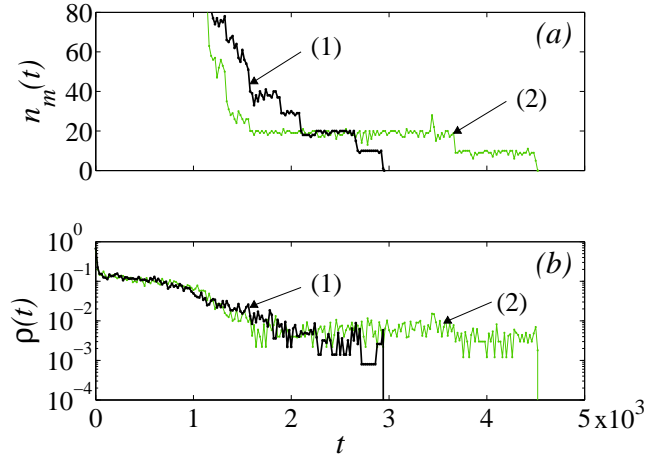}
\caption{(Color online) Time evolution of (a) the number $n_m$ of
 agents in the minority state, and (b) interface density $\rho$, for
 two typical runs that developed metastable states in an \ERhom
 network, $s=10$, \mbox{\kcoutave$=10$}, $N=1000$.}
\label{fig:metastables_ERhom}
\end{figure}

Two typical runs that developed metastable states in the ERhom network
topologies are presented in Fig.~\ref{fig:metastables_ERhom},
employing two measures: the number $n_m$ of agents in the minority
state [Fig.~\ref{fig:metastables_ERhom}(a)], and the interface density $\rho$, i.e. the fraction of
links that connect nodes in different states [Fig.~\ref{fig:metastables_ERhom}(b)]. Run (2) corresponds
to the detailed view in Fig.~\ref{fig:communityStates_ERhom_s10}. We
see that $n_m$ decreases in steps of size $s$, indicative of cliques
that are homogenized to the minority state, which are consenting to
the majority state one by one. The number of minority agents rarely
increases in the \ERhom networks. A closer inspection of the network
topologies shows that the cliques that remain in the minority state
longest have a relatively small number of out-links, although not
necessarily only one (in runs (1) and (2) displayed in
Fig.~\ref{fig:metastables_ERhom}, \kcout $=$ 4 and 6, respectively).

\begin{figure}[htb]
\centering
\includegraphics[width=1.0\linewidth]{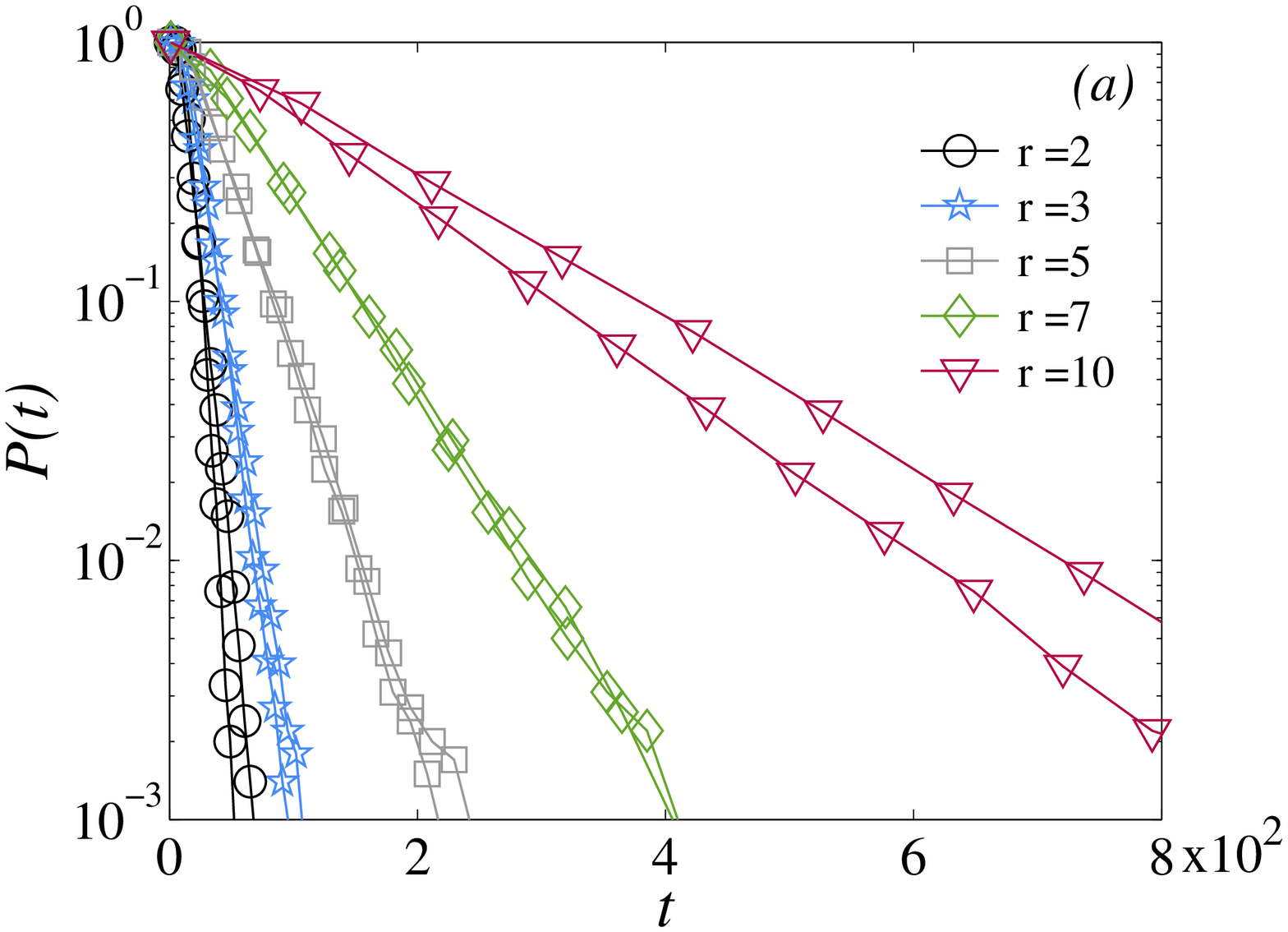}
\includegraphics[width=0.95\linewidth]{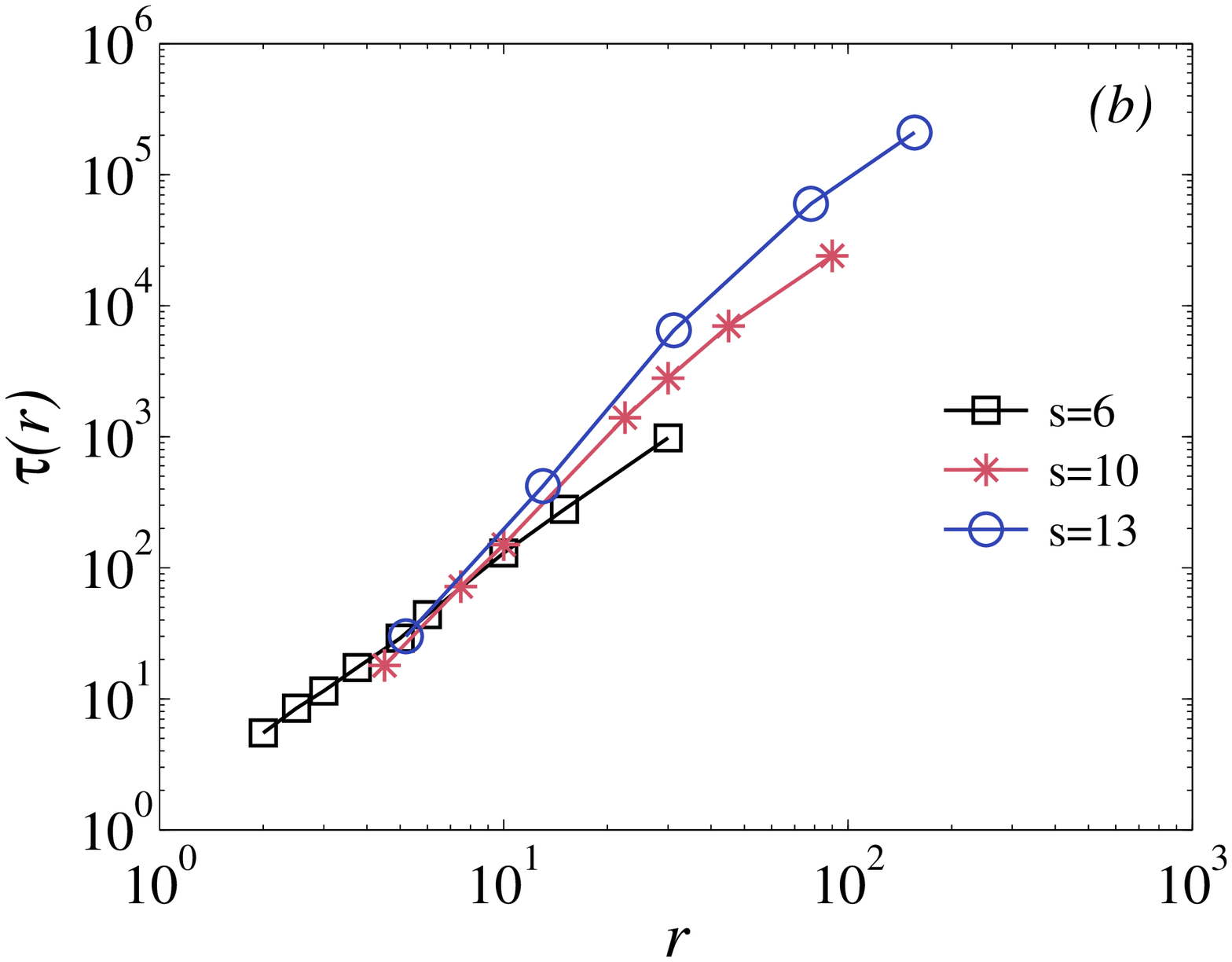}
\caption{(Color online) Dynamical robustness of cliques [as in
    Fig~\ref{schematic:dynamical robustness}(b), see text]. (a) The
fraction of alive runs $P(t) \sim e^{-t/\tau}$ shows a trend with
\mbox{\rdefsn.}  Various ratios $r$ are each represented by two pairs
of ($s,\,k_{c,out}$). From left to right: $r=2$: (6,15) and (3,3);
$r=3$: (6,10) and (4,4); $r=5$: (6,6) and (5,4); $r=7$: (8,8) and
(7,6); $r=10$: (6,3) and (10,9). We performed $10\,000$ runs in each
topology. (b) Dependence $\tau(r)$. Clique sizes $s=6$ ($\Box$),
$s=10$ ($\ast$), and $s=13$ ({\Large $\circ$}), and \kcout ranging
from $1$ to $15$, $20$, or $30$ respectively, leading to the displayed
$r$ values. }
\label{fig:invasions_fixed_r}
\end{figure}

In order to shed more light on how cliques with various out-degrees
resist changing their state, we study them in a controlled setting. In
accordance with our definition of dynamical robustness, we initialize
all agents within the clique to the state B, and the links leading out
of the clique are connected to nodes permanently in state $A$
[as shown in Fig~\ref{schematic:dynamical robustness}(b)].  As cliques remain mostly homogenized to one state
during the evolution of the dynamics, the resulting lifetime
distributions are also relevant for understanding the resistance of
communities against changing their state within the network.
Figure~\ref{fig:invasions_fixed_r}(a) displays the observed fraction
of alive runs for cliques of various sizes $s$ and out-degrees \kcoutn. The distributions are roughly exponential, $P(t)\sim
e^{-t/\tau}$, and it turns out that their survival times $\tau$ show a
clear trend with the ratio \rdefn$=s(s-1) / k_{c,out}$, which appears
to be an appropriate topological measure of the dynamical robustness
for cliques. Figure~\ref{fig:invasions_fixed_r}(b) displays the
relation $\tau(r)$, determined for cliques of fixed size with varying
clique out-degree. The time scales associated with the invasion of
cliques grow rapidly with $r$.

\begin{figure}[htb]
\centering
\includegraphics[width=0.9\linewidth]{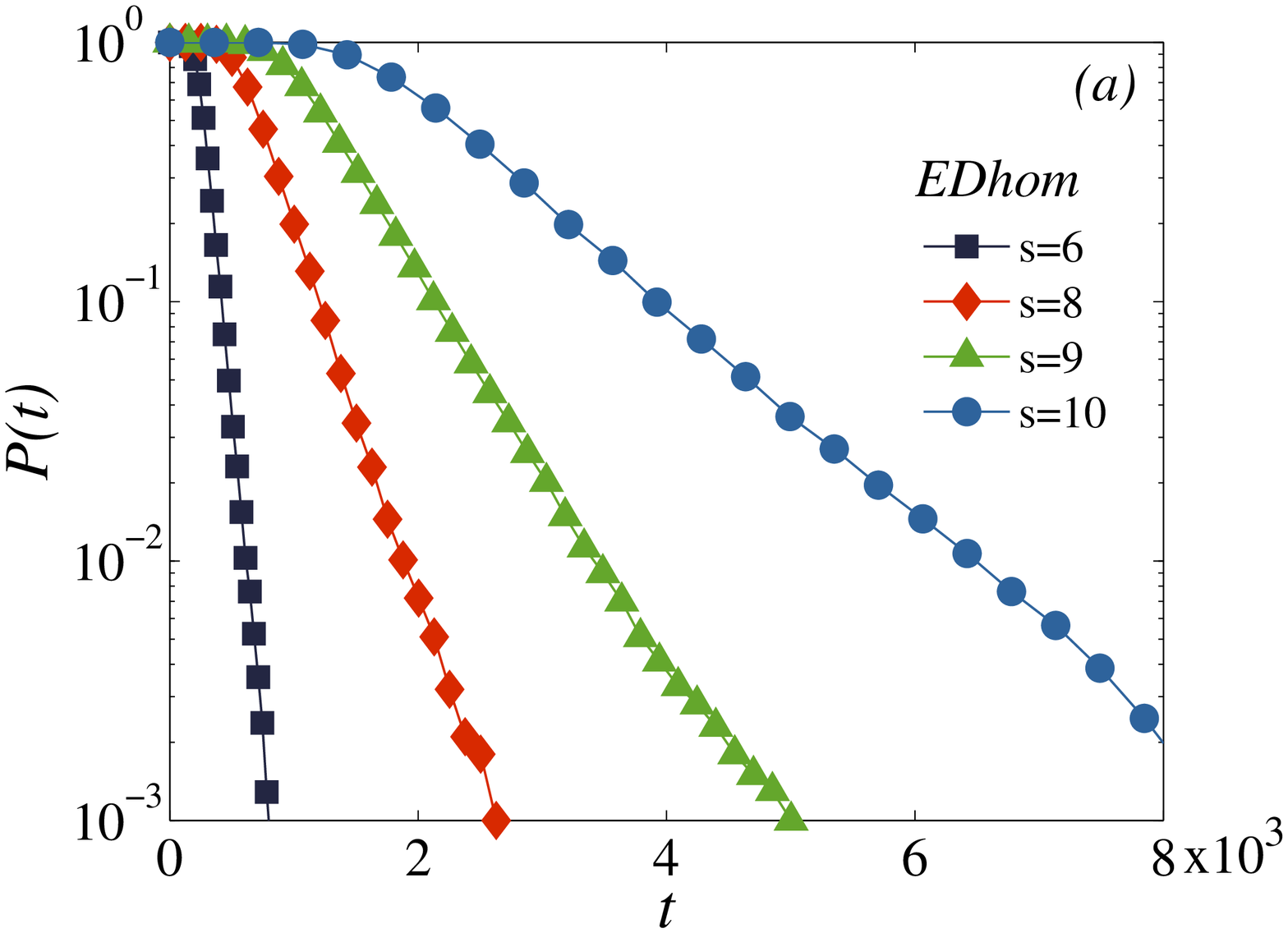}
\includegraphics[width=0.8\linewidth]{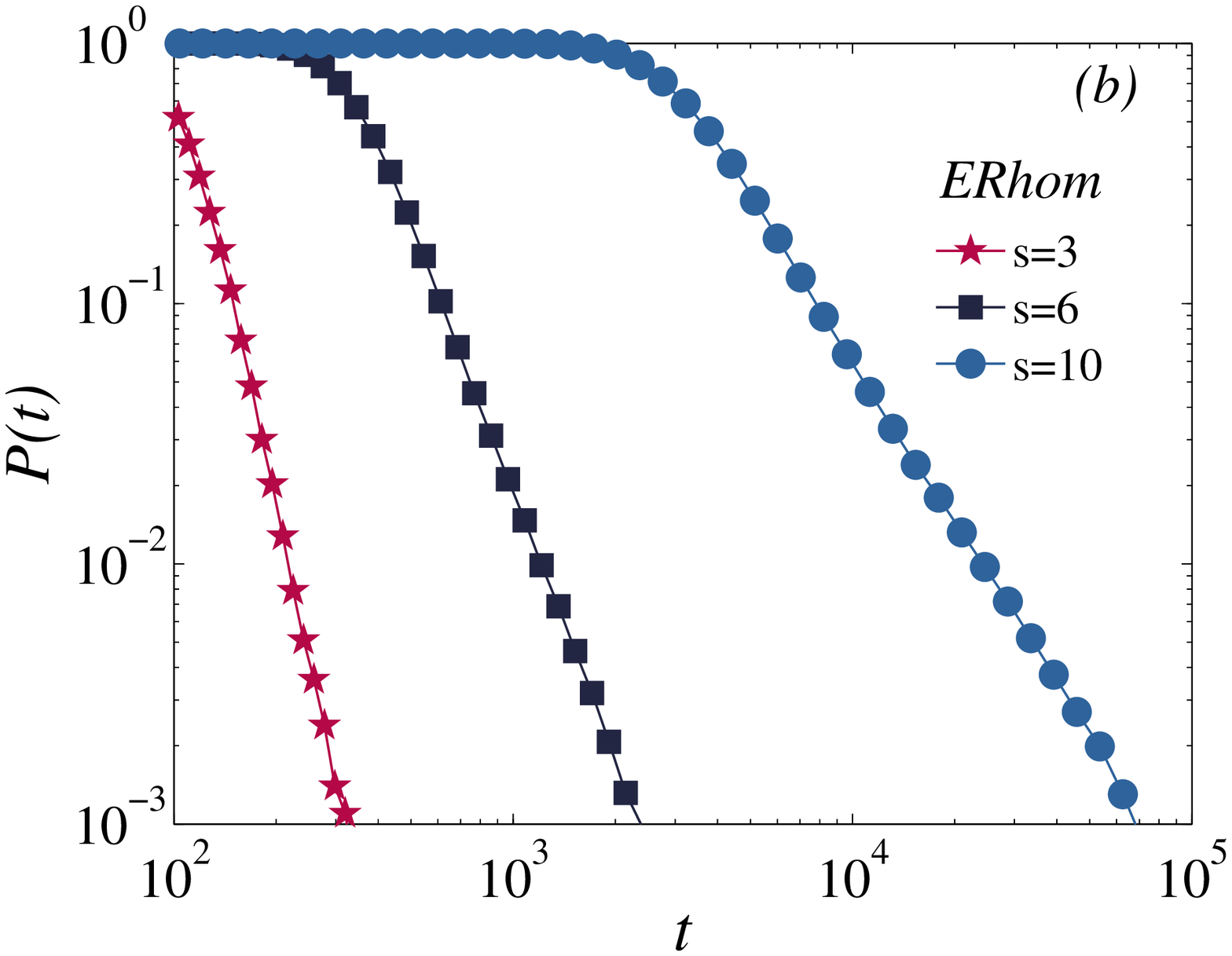}
\includegraphics[width=0.9\linewidth]{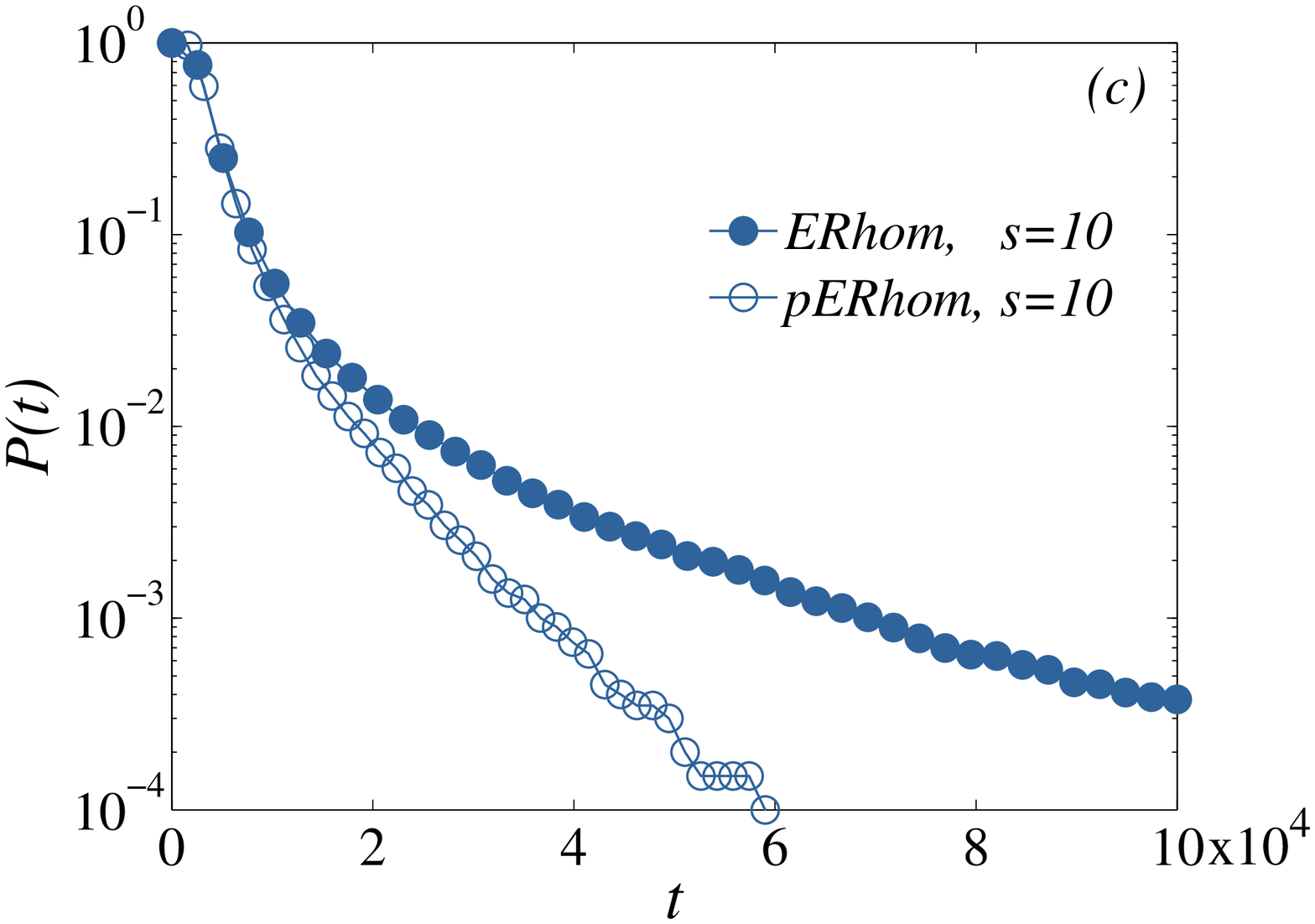}
\caption{(Color online) Fraction of alive runs in \EDhomn, \ERhomn,
  and \pERhom networks with $N\approx10^3$ and \kcout$=10$. (a) \EDhom
  with clique sizes $s=6, \, 8, \, 9,$ and $10$.  (b) \ERhom networks
  with clique sizes $s=3,6$, and $10$. (c) The \ERhom with $s=10$
  together with the corresponding \pERhom network. All cases averaged
  over 100 network realizations (except $10^3$ for \ERhom $s=6$), with
  100 runs in each.}
\label{fig:EDER}
\end{figure}

Finally, let us observe the fraction of alive runs $P(t)$ in the networks
with equally sized cliques. In \EDhom networks with clique sizes
$s=6,$ $8$, $9$, 
and $10$, and clique out-degree \kcout$=10$, $P(t)$ has an exponential
tail [Fig.~\ref{fig:EDER}(a)], indicating that the presence of
communities alone is not sufficient for a broad distribution to
appear. In these networks, all cliques have equal dynamical
robustness. In contrast, in the \ERhom networks the resulting
fractions of alive runs $P(t)$ are clearly broader than exponential,
as depicted in Fig.~\ref{fig:EDER}(b) for clique sizes $s=3,\,6,$ and
$10$, and average clique out-degree \kcoutave$=10$.  The variance in
dynamical robustness caused by the different clique out-degrees seems
to play an important role.

We probe the effect of the most isolated node groups in the \ERhom
networks by eliminating the least well connected cliques, i.e., those
that are connected to the network by a single link. This is done by
taking the two-core of the underlying ER network before replacing its
nodes with cliques. We call these networks \pERhom for ``pruned'' \ERhomn.
Figure~\ref{fig:EDER}(c) displays $P(t)$ for \ERhom and \pERhom networks
with $s=10$ and \kave$=10$. It is seen that pruning the network
results in a distribution that decays slightly faster, but remains
broader than exponential. This gives further confirmation that the
metastable states with various time scales produced by cliques with
different out-degrees are responsible for the broader than exponential
lifetimes in the \ERhom networks.

The fraction of alive runs $P(t)$ in the \ERhom networks shown in
Fig.~\ref{fig:EDER}(b) has broad tails that appear power-law
like. Moreover, they appear to broaden with increasing clique
size. Approximating these tails by power laws, the exponents would
however be far larger than those observed in the networks studied
in~\cite{comm_epl}, in which the range of exponents was such that the
variance of the lifetimes was not defined. Hence the distributions
observed here are fundamentally different from the findings
in~\cite{comm_epl}. In order to obtain broader lifetime distributions,
we apparently need a broader distribution in the dynamical robustness
of communities, which in the case of cliques can be achieved by
increasing variance in $r$. As it is more practical to obtain large
variance in $r$ by varying $s$ than \kcoutn, we take this approach in
the following section.

\subsection{Networks with a broad size distribution of clique sizes}
\label{sec:ERPL}

In this section, we study a network consisting of cliques with equal
out-degree \kcout and with an exponential clique size distribution,
shifted to obtain minimum clique size $s_{min}$.  We construct
networks from $N_c$ cliques whose sizes $s$ are obtained as $s=\lfloor
x \rfloor + s_{min}$, where $\lfloor \; \rfloor$ refers to rounding
downwards and $x$ is drawn from the exponential distribution
$p(x)=\frac{1}{\mu}exp(-x/\mu)$, leading to $p(s)\sim
\exp(-(s-s_{min})/\mu)$ for integer values of $s$ starting from
$s_{min}$. As with the \EDhom networks, the \kcout out-links of each
clique are randomly assigned to its nodes, and link ends are randomly
paired, except that no two link ends from the same clique are
connected. We label the networks as EDE, for equal out-degree
and exponential clique size distribution.

\begin{figure}[htb]
\centering
\includegraphics[width=1.0\linewidth]{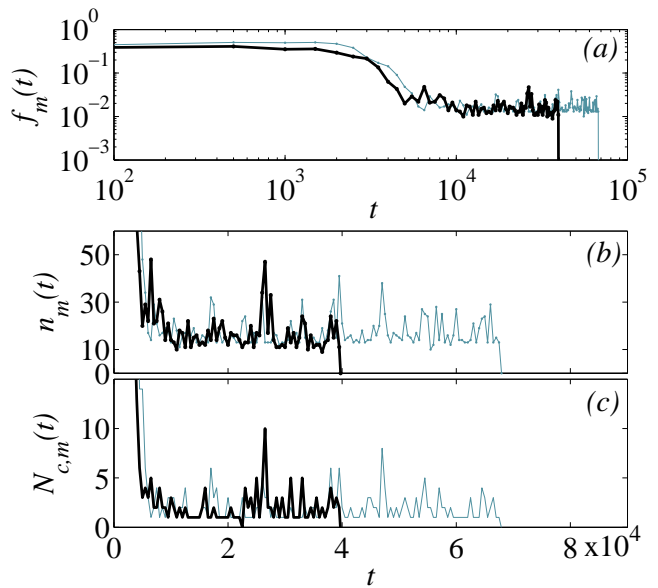}
\caption{(Color online) Two typical runs that developed metastable
  states in \EDexp networks with \kcout$=3$, $\mu=1.2$, $s_{min}=3$,
  and $N_c=270$, leading to $N\approx10^3$. Time evolution of (a) the
  fraction and (b) the number of agents in the state that became the
  minority, and (c) the number of cliques in which more than 90
  percent of agents were in the minority state.}
\label{fig:metastables_EDexp}
\end{figure}

The communities in this network will display a very large variance in
dynamical robustness. In Sec.~\ref{sec:ERhom} we saw that the factor
$\tau(r)$ of the exponential lifetime distribution of a clique being
``invaded'', grows very rapidly with $r$, which in turn grows
approximately as $r\sim s^2$.  Again, cliques remain homogenized to
either of the states $A$ or $B$ most of the time (not shown), but it turns
out that some of the smaller cliques frequently adopt the state of a
larger clique homogenized to the minority
state. Figure~\ref{fig:metastables_EDexp}(a) displays the fraction
$f_m$ of nodes in the minority state in a few typical runs that
developed metastable states in the EDE networks. A close-up of the
same runs [Fig.~\ref{fig:metastables_EDexp}(b)] shows that the number
$n_m$ of minority nodes fluctuates above a baseline of roughly
$11-13$ nodes. This seems to indicate a relatively large clique
homogenized to the minority state that is ``converting'' its smaller
(and hence less dynamically robust) neighboring cliques to the
minority state, thereby producing around itself a {\it buffer} of
cliques in the minority state. This assumption is confirmed by closer
inspection of the networks, as well as by
Fig.~\ref{fig:metastables_EDexp}(c), which shows the number $N_{c,m}$
of clusters in which more than 90 \% of the agents are in the minority
state. Much of the time there is only one cluster in the minority
[corresponding to the baseline in
Fig.~\ref{fig:metastables_EDexp}(b)], while it is frequently joined by
other, mostly smaller clusters, judging by the combination of $n_m$
and $N_{c,m}$.

The buffering effect is an additional ingredient causing metastable
states with various time scales, beyond the dynamical robustness that
depends on $r$.  We studied it in a more controlled setting using
networks consisting of a single large clique and a large number of
small cliques, with equal \kcoutn, connected similarly as in the \EDhom
networks. The $P(t)$ resulting in these networks had an exponential
tail (not shown), suggesting that buffering is again a process that
would individually have an exponential lifetime distribution, but
broader distributions are produced when a combination of substructures
with different dynamical robustness is present.

\begin{figure}[htb]
\centering
\includegraphics[width=1.0\linewidth]{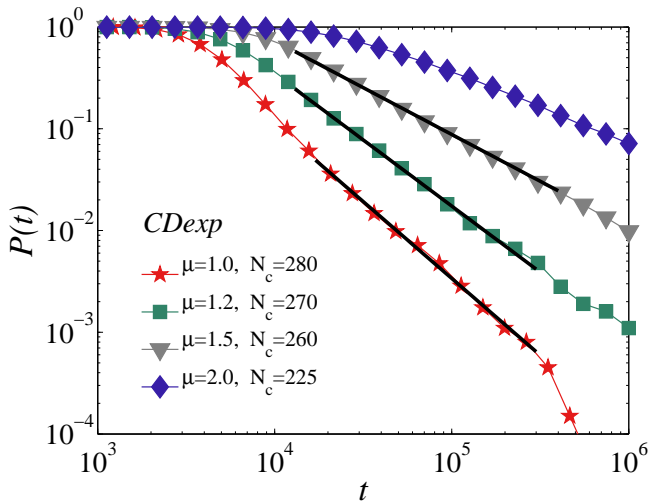}
\caption{(Color online) Fraction of alive runs $P(t)$ in \EDexp
  networks with various factors $\mu$ of the clique size distribution
  $p(s) \sim exp ( - (s-s_{min})/ \mu)$ with $s_{min}=3$. From left to
  right: $\mu=1.0, 1.2, 1.5, 2.0$ and $N_c= 280,\, 270,\, 260,\, 225$,
  leading to $N\approx1\,000$. Clique out-degree \kcout$=3$. Results
  are averaged over $1\,000$ network realizations ($2\,000$ for
  $\mu=1.0$) with $10$ runs in each. The fitted lines are power laws
  $P(t) \sim t^{- \eta} $ with exponents from left to right:
  $\eta=1.51,\,1.3,\,0.92$.}
\label{fig:aliveRuns_EDexp}
\end{figure}

The \EDexp networks are seen to give rise to very broad lifetime
distributions, shown in Fig~\ref{fig:aliveRuns_EDexp} for $s_{min}=3$,
$\mu=1.0 ... 2.0$, \kcout$=3$, and $N\approx1\,000$.  Approximating
the tails of the fraction of alive runs by a power law $P(t) \sim t^{-
\eta} $, the best fits to the cases with $\mu=1.0$ and $\mu=1.2$ have
exponents $\eta=1.51$ and $\eta=1.3$, respectively. Values $1< \eta<
2$ imply that the variance of the lifetime probability density
distribution $p(t)$ is not defined. In this way, we recover the result
found in~\cite{comm_epl}, where a characteristic lifetime could not be
defined because of the existence of trapped dynamical metastable
states. The best fit to the case with $\mu=1.5$ has an exponent
smaller than unity, $\eta=0.92$, indicating that a mean lifetime is
not defined either. We note that for each network realization, the
community sizes are sampled from a distribution, and the observed
broad lifetime distribution is a result of averaging over several runs
in many network realizations.

\section{Conclusions}
\label{sec:discussion}

In this study we set out to determine minimal network features that
would produce broad lifetime distributions for the ordering dynamics
described by the AB model.  We have introduced the concept of
dynamical robustness against invasion in relation to the dynamics of
competing options to describe the resistance against outside influence
of topological substructures that involve relative isolation from the
rest of the network. Dynamical robustness is characterized by the
survival time of the substructure, i.e., the characteristic time needed for this
set of nodes before changing its option toward the one of the
surrounding majority.  In all of the topologies in which a broader
than exponential distribution for the relaxation time of the whole
system arose, we have identified substructures that individually have
exponential lifetime distributions, implying a well defined survival
time for such topologies.  The broad distribution appears because of
the heterogeneity of these substructures, such that the network has a
variety of different substructures with different survival times.

In an \er network, branches were seen to produce exponential lifetime
distributions when isolated from the rest of the network. Their
dynamical robustness has been proven to be affected by the diameter as
well as the number and location of bifurcations.  Lifetime
distributions also appear to be exponential for isolated cliques, and
the ratio \rdef has proven an appropriate topological measure to
characterize the dynamical robustness of a clique.

In the case of networks with mesoscale structure built up from
randomly connected cliques, it has been seen that simply the presence
of communities is not a sufficient condition to produce a broader than
exponential lifetime distribution. This was demonstrated by networks
consisting of cliques with equal size and same out-degree, and hence
equal dynamical robustness (\EDhomn), where the lifetime distribution
for the whole network has proven to be exponential.  Although the
interactions between the cliques in a network may cause clique
lifetimes to deviate from those that arise in isolation, the broader
than exponential lifetime distributions observed for \ERhom and \EDexp
may in part be explained by the different dynamical robustness against
invasion of the cliques forming the network, leading to a combination
of exponential processes with various time scales. The most
interesting feature is obtained for \EDexp networks where we have
recovered the main results in~\cite{comm_epl}, i.e. very broad $P(t)$
with a best power law fit such that the second moment of the
distribution is not defined, and therefore there does not exist a
characteristic time scale for the dynamics. The results in this paper
might be generic for a class of models where the dynamics at the
interfaces is curvature driven.

In summary, complementing studies on the effects of heterogeneous
interacting agents (a research line of growing
interest~\cite{Tessone_2006}), we have seen that heterogeneity at the
mesoscale level of the network of interaction results in non-trivial
effects in the dynamics of ordering processes. A large variability in
the dynamical robustness of different topological substructures
appears to be a sufficient mechanism for the absence of a
characteristic time for the dynamics. This mechanism causes the
existence of dynamical metastable states that survive at any time
scale. Substructures might have a well defined survival time, but the
existence of a variety of substructures with different dynamical
robustness (characterized by different survival times) results in a
very broad $P(t)$ that does not allow to identify a characteristic
relaxation time for the whole system.

\section*{Acknowledgments}

The authors acknowledge financial support from EU COST action P10 that
enabled co-operation in the form of research visits. The work was also
supported by the MEC (Spain) through project FISICOS (FIS2007-60327),
by the European Commission through the NESTComplexity project PATRES
(043268), and by the Academy of Finland, Center of Excellence program
2006-2011.  X.C. also acknowledges financial support from a
Ph.D. fellowship of the Govern de les Illes Balears (Spain), and
R.T. was also supported by the ComMIT graduate school.


\end{document}